\documentclass[a4paper,]{article} 
\usepackage[top=15mm, bottom=15mm, left=15mm, right=15mm]{geometry}
\usepackage{fullwidth}
\usepackage{graphicx} 
\usepackage{epsfig} 
\usepackage{mathptmx} 
\usepackage{times} 
\usepackage{amsmath} 
\usepackage{amssymb}  

\usepackage{algpseudocode} 
\usepackage{units}
\usepackage{color}
\usepackage{url}

\author{Vito Trianni$^{1}$, Daniele De Simone$^{2}$, Andreagiovanni
  Reina$^{3}$ and Andrea Baronchelli$^{4}$  \vspace{2ex}\\
  \small$^{1}$ISTC, National Research Council, 00185 Rome, Italy\\
  {\tt\small vito.trianni@istc.cnr.it}\\
  \small$^{2}$DIAG, Sapienza University of Rome, 00185 Rome, Italy\\
  {\tt\small desimone@diag.uniroma1.it}\\
  \small$^{3}$University of Sheffield, Sheffield S1 4DP, UK\\
  {\tt\small a.reina@sheffield.ac.uk}\\
  \small$^{4}$City University London, London EC1V 0HB, UK\\
  {\tt\small  a.baronchelli.work@gmail.com}}

\title{Emergence of Consensus in a Multi-Robot Network:\\from Abstract Models to Empirical Validation}
\date{}

\begin{document}

\maketitle

\begin{abstract}
  Consensus dynamics in decentralised multiagent systems are subject
  to intense studies, and several different models have been proposed
  and analysed. Among these, the \emph{naming game} stands out for its
  simplicity and applicability to a wide range of phenomena and
  applications, from semiotics to engineering. Despite the wide range
  of studies available, the implementation of theoretical models in
  real distributed systems is not always straightforward, as the
  physical platform imposes several constraints that may have a
  bearing on the consensus dynamics. In this paper, we investigate the
  effects of an implementation of the naming game for the
  \emph{kilobot} robotic platform, in which we consider concurrent
  execution of games and physical interferences. Consensus dynamics
  are analysed in the light of the continuously evolving communication
  network created by the robots, highlighting how the different
  regimes crucially depend on the robot density and on their ability
  to spread widely in the experimental arena. We find that physical
  interferences reduce the benefits resulting from robot mobility in
  terms of consensus time, but also result in lower cognitive load for
  individual agents.
\end{abstract}

\section{Introduction}
\label{sec:introduction}

Collective decision-making is an essential capability of large-scale
decentralised systems like robot swarms, and is often key to achieve
the desired goal. In swarm robotics, a large number of robots
coordinate and cooperate to solve a problem, and often consensus among
the robots is necessary to maximise the system performance
\cite{Parker:TraMech:2009,Sartoretti:2014et,Reina:2015hu}. The design
of controllers for consensus decision is often inspired by models of
collective behaviour derived from studies in the ethology of social
systems \cite{Seeley:Science:2012,Reid:2015gi}, as well as from
studies about the emergence of social conventions and cultural traits
\cite{Sood:2005fr,Castellano:2009ce,Brockmann:2013bq}.
Theoretical models represent idealised instances of collective
decentralised systems in which consensus can be somehow attained.
Among the different available models, a particularly interesting case
is the one of the \emph{naming game} (NG), which represents the
emergence of conventions in social systems, such as linguistic,
cultural, or economic conventions
\cite{steels1995self,Baronchelli:2006tl,Centola:2015eo}. The appeal of
this model consists in the ability to describe the emergence of
consensus out of a virtually infinite set of equivalent alternatives,
yet requiring minimal cognitive load from the agents composing the
system \cite{Baronchelli:2006tl,Baronchelli:2006ft}.  Moreover, the NG
has been successfully demonstrated on a network of mobile point-size
agents \cite{Baronchelli:2012dq}. Such a collective decision-making
behaviour can be very useful in swarm robotics in case consensus is
required with respect to a possibly large number of alternatives
(e.g., the location and structure for cooperative construction
\cite{Werfel:2014ic,Soleymani:2015ras}, or the most functional shape
for self-assembly \cite{OGrady:2009di,Rubenstein:2014dt}).

When dealing with the implementation of physical systems starting from
theoretical models, however, several constraints may arise which may
have a bearing on the collective dynamics. Indeed, small
implementation details at the microscopic scale may have a large
impact at the macroscopic level. Hence, it is important to study the
effects of such constraints in relation to the dynamics predicted by
the theoretical models.
In this paper, we propose an implementation of the NG for the
\emph{kilobot} robotic platform \cite{Rubenstein:2013dq}.  Kilobots
are low-cost autonomous robots designed for experimentation with large
groups \cite{Rubenstein:2014dt}. They can move on a flat surface and
interact with close neighbours by exchanging short messages sent on an
infrared channel.  The collective behaviour of a kilobot swarm results
solely from the individual decisions and inter-individual
interactions, without any central unit directing the group dynamics.
As a consequence, the implementation of the NG for the kilobots needs
to be fully decentralised with games autonomously triggered by any
robot at any time.
Additionally, within a decentralised system, the concurrent execution
of games by neighbouring robots becomes possible, in opposition to the
rigorously sequential scheme typically adopted in theoretical
studies. Hence, the interaction pattern among robots may be
significantly altered, and the corresponding dynamics need to be
carefully characterised.
%
Finally, the embodiment of the robots determines physical
interferences (i.e., collisions) that strongly influence the overall
mobility pattern. It follows that abstract models of agent mobility
must be contrasted with experimentation with robots, in which all the
details of the physical platform can be taken into account.

In this paper, we study the effects of the motion and interaction
patterns on the consensus dynamics, and we pay particular attention to
both concurrent executions of games and physical interferences.
First, we provide an abstract model of mobile agents playing the NG,
in which physical interferences are ignored. Following previous
studies \cite{Baronchelli:2012dq}, we analyse this model in the light
of the communication network established by the agents, we show how
the consensus dynamics are determined by agent density, mobility and
interaction frequency, and we link our empirical findings with
theoretical studies \cite{Baronchelli:2006ft,Lu:2008fq}. Then, we
contrast abstract models with large-scale simulations of the kilobots,
as well as with real-world experiments. Here, physical interferences
impact on the consensus dynamics by limiting the free diffusion of
robots in the experimental arena, hence increasing consensus
times. Still, the cognitive load for the individual agents is reduced
for physical implementations, due to the lower number of alternatives
that each agent must consider in average.
Our adapted implementation of the NG is presented in
Section~\ref{sec:model}, while the corresponding consensus dynamics
are discussed in Section~\ref{sec:consensus-dynamics}. Conclusions and
future directions are presented in Section~\ref{sec:conclusions}.

\section{Model and Implementations}
\label{sec:model}

The naming game in its basic form \cite{Baronchelli:2006tl} models
pairwise interactions in which two players---the speaker and the
hearer---interact by exchanging a single word chosen by the speaker,
and updating their inventory on the basis of the game
success. Previous extensions of the model take into account different
inventory updating and communication schemes \cite{Baronchelli:2011ka}
and also consider mobile agents \cite{Baronchelli:2012dq}. In this
work, we adopt a broadcasting scheme for the speaker agent, while
inventory updating is performed only by the hearer agent, as detailed
in the following (for details, see \cite{Baronchelli:2011ka}).

When engaging in a NG, the speaker agent~$a_s$ selects a word $w$
either randomly from its inventory, or inventing it anew should the
inventory be empty (i.e., the set of possible choices for a new word
$w$ is virtually infinite). Then, it broadcasts $w$ to all agents in
its neighbourhood. Upon reception of $w$, the hearer agent $a_h$
updates its inventory by either storing $w$ if it was not found in
$a_h$'s inventory, or by removing all words but $w$ if the latter was
already known to $a_h$. By iterating the game multiple times, the
entire system converges toward the selection of a single word shared
by all agents \cite{Baronchelli:2006tl,Baronchelli:2011ka}.

\subsection{Multiagent simulations}
\label{sec:ma-simul}

We implement a decentralised version of the NG by letting each agent
$a$ autonomously take the role of speaker every
$\unit[\tau_s]{s}$. Given that agents update their state at discrete
steps of $\delta_t=\unit[0.1]{s}$, they communicate every $n_s$ steps
so that a word is broadcast every $\unit[\tau_s= n_s\delta_t]{s}$ to
all neighbours within the range $d_i=\unit[10]{cm}$. In this way,
concurrent execution of games becomes possible, hence introducing an
important difference from previous theoretical studies in which at any
time only one game is executed by a randomly chosen agent and one of
its neighbours
\cite{Baronchelli:2012dq,Baronchelli:2006tl,Baronchelli:2011ka,Lu:2008fq}.
At the hearer side, multiple interactions are possible within any time
interval $\delta_t$, depending on the local density of agents. Hence,
all words received in a single $\delta_t$ period are used sequentially
to update the inventory. The list of received words is randomised
before usage to account for the asynchronous reception of messages by
the physical platform (see
Section~\ref{sec:kilobot-implementation}). %
The algorithm for the multiagent implementation is shown in
Fig.~\ref{fig:mng-algorithm}.

At the beginning of each simulation run, $N$ agents are deployed
uniformly random within a squared box of side $L$ with periodic
boundary conditions (e.g., a torus). This allows to focus on the
effects of agent mobility and density without constraints from a
bounded space \cite{Baronchelli:2012dq}. Agents are dimensionless
particles and therefore do not collide with each other. The agents
neighbourhood is determined by all the agents within the interaction
range $d_i$. By moving in space, the agent neighbourhood varies so
that a dynamic communication network is formed.  Agent mobility
follows an uncorrelated random walk scheme, with constant speed
$v=\unit[1]{cm/s}$ and fixed step length $v\tau_m$, where $\tau_m$
represents the constant time interval between two consecutive
uncorrelated changes of motion direction. This leads to a diffusive
motion with coefficient $D\sim\,v^2\tau_m$
\cite{Baronchelli:2012dq}. In practice, agents change direction every
$n_m$ steps, so that $\tau_m= n_m\delta_t$ (see
Fig.~\ref{fig:mng-algorithm}).

\begin{figure}[t] 
  \begin{fullwidth}[width=0.5 \linewidth]
    \begin{algorithmic}[H] \small
      \Procedure{NG}{$n_m$, $n_s$}\Comment{Implementation of the NG}
      \State $n_t\gets n_t + 1$
      \If{$n_t\mod n_m = 0$}\Comment{Change motion direction} \State
      RandomTurn()
      \EndIf
      \State MoveStraight()
      
      \State $\mathbf{W}\gets$ReceiveWords() \Comment{Play the hearer role}
      \State Randomise($\mathbf{W}$)
      \For{$w\in\mathbf{W}$}
      \State UpdateInventory($w$)
      \EndFor
      
      \If{$n_t\mod n_s = 0$}    \Comment{Play the speaker role} 
      \State $w \gets $ SelectWord()
      \State Broadcast($w$)
      \EndIf
      
      \EndProcedure  
    \end{algorithmic}
  \end{fullwidth}
  \caption{The NG algorithm exploited in multi-agent simulations.}
  \label{fig:mng-algorithm}  
\end{figure}

\subsection{Robot simulations}
\label{sec:robot-simulations}

We have developed a custom plugin for simulating kilobots within the
ARGoS framework \cite{Pinciroli:2012dc}, paying particular attention
to match the real robot features in terms of body size, motion speed
and communication range. Communication is implemented by allowing the
exchange of messages between neighbours within the range $d_i$. No
failures in communication have been simulated, assuming that the
channel can support communication even with high densities. We will
discuss this choice in the light of the obtained results in
Section~\ref{sec:conclusions}.

Concerning the motion pattern, kilobots are limited to three modes of
motion: forward motion when both left and right motors are activated,
and left or right turns when only one motor is activated. Turning is
performed while pivoting on one of the kilobot legs. We have therefore
implemented a differential drive motion scheme centred between the two
backward legs of the kilobot, with speed $v = \unit[1]{cm/s}$ for
forward motion and angular speed $\omega = \unit[\pi/5]{s^{-1}}$ for
turning. A multiplicative gaussian noise applied at every simulation
cycle (standard deviation $\sigma=0.4$) simulates the imprecise motion
of kilobots.  With such an implementation, the individual motion is
still diffusive, but with a lower coefficient due to the delay
introduced by turning. Additionally, collision avoidance is not
possible with the kilobot onboard sensors, and robots are let free to
crash into walls and each other. The ARGoS framework provides a 2D
dynamics physics engine that handles collisions between robots and with
walls with an integration step size $\delta_t=\unit[0.1]{s}$, which
proves sufficient for our purposes. Collisions determine a further
reduction in the diffusion speed, as we will discuss in
Section~\ref{sec:robot-results}.
Robots are deployed randomly within a squared box of side $L$
surrounded by walls. To avoid overlapping of robots, the initial
positions are determined by dividing the arena in cells wide enough to
contain a single kilobot, and randomly placing kilobots into free
cells.

\subsection{Kilobot implementation}
\label{sec:kilobot-implementation}

The implementation of the NG for kilobots requires handling
transmission and reception of messages, and implementing the random
walk. We use the kilobot API from Kilobotics \cite{kilobotics}, which
provides two callback functions for transmission and reception of
10-byte messages, functions for distance estimation of the message
source, and a counter that is updated approximately 32 times per
second (i.e., $\delta_t\simeq 1/32$). Broadcast is allowed every
$\tau_s$ seconds by opportunely activating the transmission
callback. Communication interferences among robots are treated through
the CSMA-CD protocol (carrier-sensing multiple access with collision
detection) with exponential back-off, meaning that upon detection of
the occupied channel, message sending is delayed within an
exponentially increasing range of time slots. This introduces an
additional level of asynchrony that must be tolerated by the
collective decision-making process, as the exact timing of
communication cannot be completely controlled.
Upon reception of any message, the corresponding callback function is
activated, and the NG is immediately played exploiting the content of
the received message. Given that the maximum communication distance
may vary across different robots, we capped the maximum distance to
$d_i$ by software, estimating the source distance and ignoring
messages from sources farther than $d_i$.

The motion pattern implements the random walk exactly as performed in
simulation, exploiting the internal random number generator for
uniformly distributed turning angles. Forward motion $v$ and angular
speed $\omega$ of each kilobot have been calibrated to obtain a
roughly constant behaviour across different robots and to match the
parameter values used in simulation. The code for the controller is
written in a \texttt{C}-like language (\texttt{AVR C}) and fits in
about 200 lines.
In experimental runs, kilobots are initially positioned randomly
following indications from the ARGoS simulator in equivalent
conditions. This provides an unbiased initialisation and supports
comparison with simulations in Section~\ref{sec:validation-with-real}.


\section{Consensus Dynamics}
\label{sec:consensus-dynamics}

\newcommand{\ak}{\langle k\rangle}

The most important quantity to evaluate the consensus dynamics
following the NG process is the time of convergence $t_c$, i.e., the
time required for the entire group to achieve consensus. Previous
studies demonstrated that consensus is the only possible outcome, even
though in particular cases it can be reached only asymptotically
\cite{Baronchelli:2006tl}.
Another relevant metric for the NG in multiagent systems is the
maximum memory $M$ required for the agents, in average, until
convergence: given that each agent needs to store a possibly large
number of words, it is important to study how the memory requirements
scale with the system size, especially in the perspective of the
implementation for real robots that entail limited memory and minimal
processing power to search large inventories.

Following previous studies, it is useful to look at the (static)
interaction network resulting by linking all agents that are within
interaction range. Given $N$ agents confined in a $L\times L$ space
and interacting over a range $d_i$, the resulting network has average
degree $\ak = \pi N d_i^2/L^2$
\cite{Fujiwara:2011fu,Baronchelli:2012dq}. Given that in our case all
parameters are constant but the agent density (as determined by $N$),
two values are critical:
\begin{enumerate}
\item $N_1 = N_{\ak=1}$ is the group size at which the average degree
  is around 1, meaning that each agent has in average one other agent
  to interact with. Below this value, interactions are sporadic and
  determined by the agent mobility, while above this value
  interactions are frequent as small clusters of agents appear.
\item $N_c = N_{\ak\simeq 4.51}$ corresponds to the critical group
  size for a percolation transition \cite{Fujiwara:2011fu}. Above
  $N_c$, the network is characterised by a giant component of size
  $N$.
\end{enumerate}
Given the broadcasting rule employed for the NG in this paper, it is
clear that the characteristics of the interaction network are
fundamental. If there exists a giant component, information can spread
quickly. If otherwise robots are mostly isolated, they will not be
able to interact and convergence would be slower, as discussed in the
following.



\subsection{Influence of density, mobility and interaction frequency}
\label{sec:ma-results}

\begin{figure}[!t]
  \centering
  \includegraphics[width=0.6\textwidth]{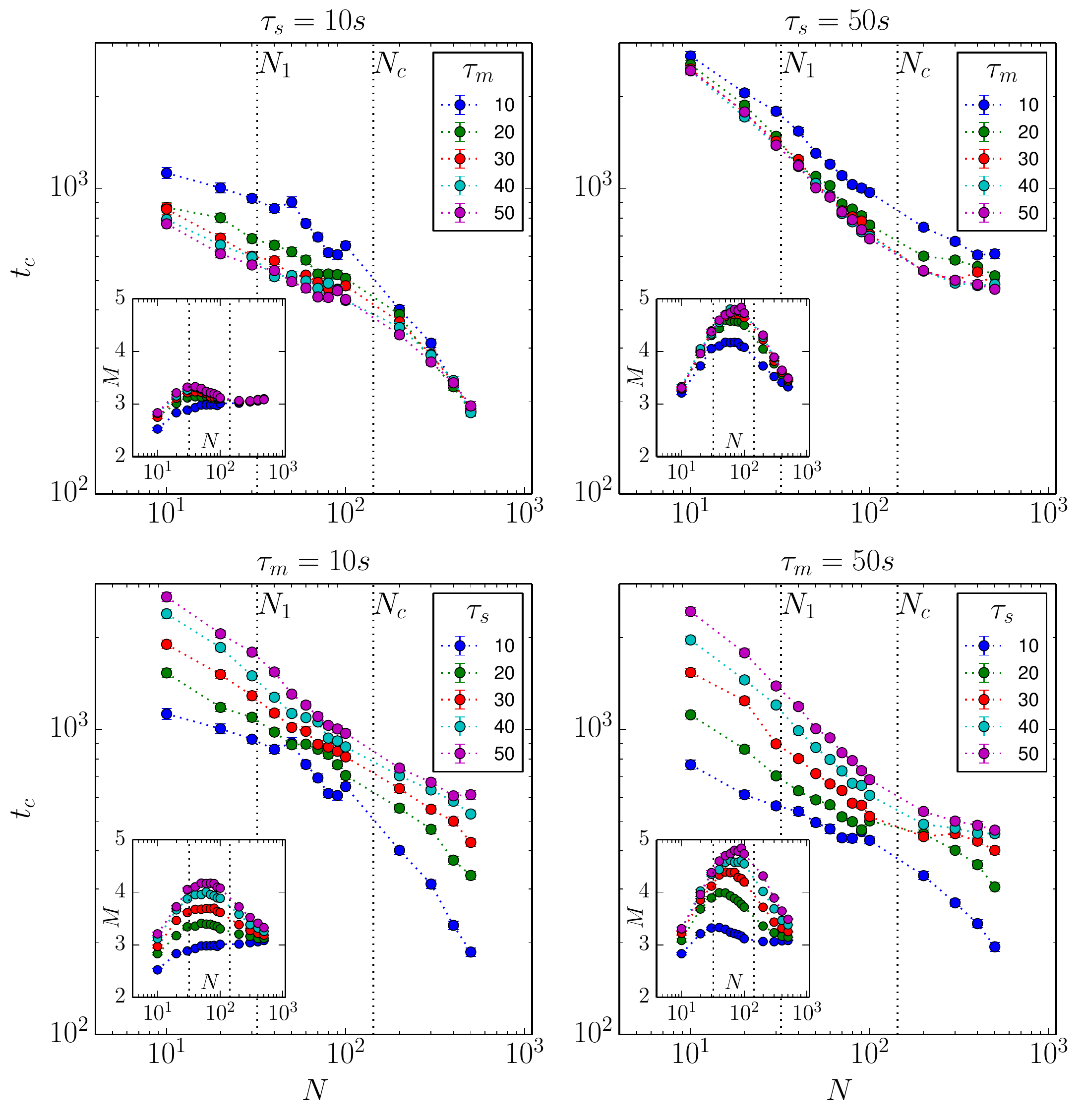}
  \caption{Results from multiagent simulations. Each panel shows the
    dependence of the convergence time $t_c$ on the system size $N$
    for different parameterisation. Statistical error bars are not
    visible on the scale of the graphs. Vertical dotted lines indicate
    the thresholds for $N_1=32$ and $N_c=143$. The insets show the
    memory requirements $M$ plotted against the system size $N$ for
    the same parameterisations.}
  \label{fig:multiagent-simulations}
\end{figure}

To determine the consensus dynamics and the effects of the different
parameters of the system, we run multiagent simulations with small and
large groups in a squared arena of size $L=\unit[1]{m}$. In this
condition, we have $N_1\simeq32$ and $N_c\simeq143$. Figure
\ref{fig:multiagent-simulations} reports the consensus time $t_c$ for
different parameterisations varying $N \in [10,500]$ and
$\tau_m,\tau_s\in \unit[[10,50]]{s}$.

\begin{figure}[!t]
  \centering
  \includegraphics[width=0.6\textwidth]{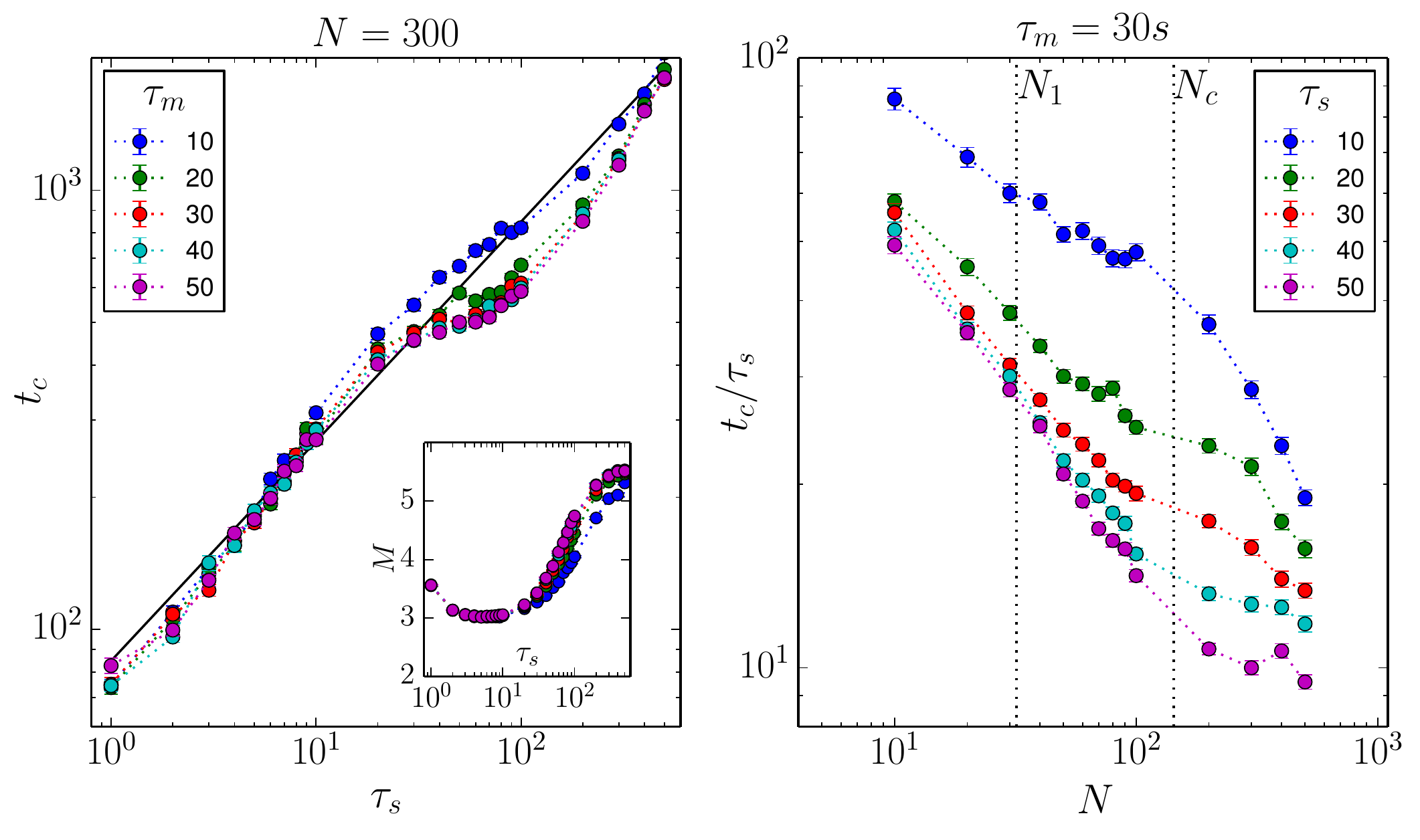}
  \caption{(Left) Scaling of convergence time as a funciton of
    $\tau_s$. The black solid line $t_c \simeq \tau_s^{0.5}$ serves as
    a guide for the eye to appreciate the power law scaling of the
    convergence time $t_c$. (Right) Rescaling convergence time by
    $\tau_s$, representing the average number of broadcasts before
    convergence. Statistical error bars are not visible on the scale
    of the graphs.}
  \label{fig:rescaling}
\end{figure}

Looking at the results, we note that $t_c$ is a decreasing function of
$N$. Indeed, a higher density corresponds to a higher number of
concurrently executed games, and results in a faster convergence.  For
$N>N_c$ and small $\tau_s$, the consensus time collapses to the same
value for varying $\tau_m$ (see for instance the top-left panel in
Fig.~\ref{fig:multiagent-simulations}). Above the percolation
threshold, agent mobility plays a minor role and the consensus
dynamics can be related to the characteristics of the static network
of interactions. Especially for low values of $\tau_m$, the dynamics
closely correspond to those of static agents interacting on a random
geometric network \cite{Lu:2008fq}. Here, the agreement process
proceeds through the formation of clusters of agents with local
consensus separated by an interface of ``undecided'' agents, and
consensus dynamics recall the coarsening on regular lattices
\cite{Baronchelli:2006ft}. This is confirmed by the left panel in
Fig.~\ref{fig:rescaling}, which shows how the convergence time $t_c$
scales with $\tau_s\in\unit[[1,500]]{s}$ for $N=300$. It is possible
to appreciate a kind of power-law scaling $t_c \simeq \tau_s^\gamma$,
with $\gamma\simeq 0.5$. This indicates that the convergence dynamics
are mostly determined by $\tau_s$, while $\tau_m$ plays a relatively
minor role, hence confirming the above mentioned resemblance with
coarsening on lattices or random geometric networks. Similarly to
fully-connected networks \cite{Baronchelli:2006tl}, log-periodic
oscillations are visible in the power law scaling, so that for some
values of $\tau_s$, mobility happens to be more relevant, with large
$\tau_m$ determining a lower convergence time (see also
Fig.~\ref{fig:multiagent-simulations} top-right).


Below the percolation threshold $N_c$, agents form temporary clusters
that dissolve due to the agent mobility. If the density is still high
enough to ensure frequent interactions ($N>N_1$), the dynamics are
determined more by the mobility of agents than by the broadcasting
period $\tau_s$. This is visible in the bottom-left and bottom-right
panels of Fig.~\ref{fig:multiagent-simulations}, where convergence
times tend to coalesce for different values of $\tau_s$ and
$N\in[N_1,N_c]$. Instead, for very low densities ($N<N_1$),
agent-agent contacts are infrequent and last for short periods of
time, so that many broadcasts go unnoticed. In this condition, high
mobility is important as much as short broadcasting periods to ensure
faster convergence (see Fig.~\ref{fig:multiagent-simulations}).

To evaluate the effects of the broadcasting period more thoroughly, it
is useful to look at the rescaled time $t_c/\tau_s$ indicating the
average number of broadcasts each agent transmitted (see
Fig.~\ref{fig:rescaling} right). We note lower values of the rescaled
agreement time for larger values of $\tau_s$, meaning that the number
of broadcasts required for convergence diminishes for longer
broadcasting periods, recalling the slower-is-faster effect observed
in many complex systems \cite{Gershenson:2015ez}.
%

A look at the memory requirements reveals that $M$ is generally
constrained to low values, which makes the NG implementation
affordable for physical systems (see the insets in
Fig.~\ref{fig:multiagent-simulations}).
For $N_1<N<N_c$, mobility plays a significant role, with larger values
of $\tau_m$ corresponding to larger $M$. The transient formation of
small clusters enhances the requirements of memory the more the agents
are able to travel between clusters that agree on different
words. Similarly, if we look at the bottom panels, we notice that
higher values of $\tau_s$ determine higher values for $M$. Here, slow
convergence leads to agents diffusing in the arena and being exposed
to multiple options, hence increasing the memory requirements.  For
$N>N_c$, instead, mobility is less important and the memory
requirements are bound to the interaction period. The scaling analysis
presented in the left panel of Fig.~\ref{fig:rescaling} shows that the
memory requirements increase drastically with $\tau_s$, confirming
that slower convergence implies also larger memory requirements. On
the other hand, frequent interactions lead to the quick formation of
few clusters, so that the individual memory requirements are limited
to few words, especially for those agents at the interface between
clusters. As $\tau_s$ decreases further, the effect of concurrent
executions of games starts to be visible with an increase in the
memory requirements as a result of the higher probability of agents
  to simultaneously exchange different words.

%


 

\begin{figure}[!t]
  \centering
  \includegraphics[width=0.6\textwidth]{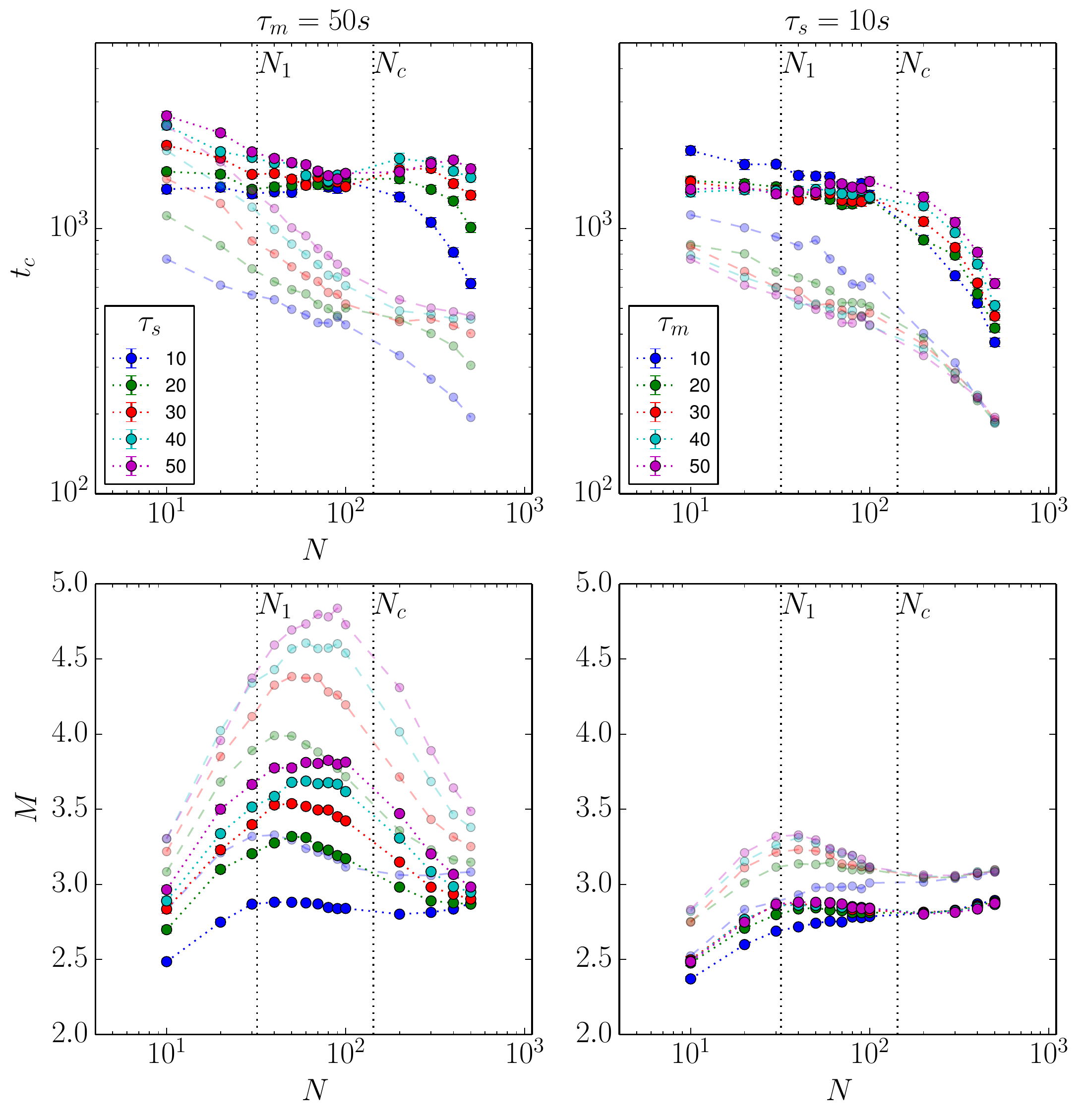}  
  \caption{Comparison between multiagent and robot
    simulations. Average values from multiagent simulations are shown
    with transparent colours, and serve as reference to appreciate the
    results from robotic simulations. Statistical error bars are not
    visible on the scale of the graphs. (Top) Convergence time
    $t_c$. (Bottom) Required memory $M$. (Left) Results for
    $\tau_m=\unit[50]{s}$, and varying $\tau_s$. (Right) Results for
    $\tau_s=\unit[10]{s}$, and varying $\tau_m$.}
  \label{fig:comparison-sim}
\end{figure}

\subsection{Influence of physical interferences}
\label{sec:robot-results}

The consensus dynamics described above refer to an ideal system that
neglects the physical embodiment of robots. Embodiment leads to
collisions with walls and among robots that constrain mobility. We
study the influence of embodiment by comparing multiagent with
robotics simulations performed in similar conditions to what described
above (see Fig.~\ref{fig:comparison-sim} for selected
parameterisations). We first note that the convergence time $t_c$ is
in general higher for robotic simulations, as a consequence of the
slower diffusion in space resulting from the turning time, which
introduces a stochastic delay in the random walk pattern, and due to
the physical boundaries that prevent robots to freely move. For
instance, in the top-left panel of Fig.~\ref{fig:comparison-sim} we
show the case for $\tau_m=\unit[50]{s}$: here, collisions lead to an
approximately constant $t_c$ for $N_1<N<N_c$, no matter what is the
broadcasting time $\tau_s$. Indeed, the slower diffusion and the
formation of small clusters determine the convergence time more than
the interaction frequency.
Collisions with walls and with other robots lead to the formation of
stable clusters in which consensus can be quickly achieved. Such
clusters dissolve at a slower pace for larger values of $\tau_m$, due
to robots turning away less often. Hence, the effects of mobility are
diluted especially when it is supposed to play an important role,
i.e., when $N<N_c$. Collisions influence the convergence dynamics also
for $N>N_c$, although to a lesser extent, as can be seen in the
top-right panel in Fig.~\ref{fig:comparison-sim}: for large $\tau_m$,
convergence is slower due to the formation of clusters that do not
interact frequently, as collisions prevent robots to mix as much as in
the ideal multiagent case.

The low ability to mix due to collisions has an effect also on the
required memory $M$, which is in general lower for robot simulations
(see bottom panels of Fig.~\ref{fig:comparison-sim}). The slower
diffusion of robots in space and the existence of boundaries limit the
spreading of different words into the robot network, hence resulting
in lower memory requirements.

\begin{figure}[!b]
  \centering\scriptsize
  \begin{tabular}{c@{\quad}c@{\quad}c@{\quad}c@{\quad}c}
    \includegraphics[width=0.18\textwidth]{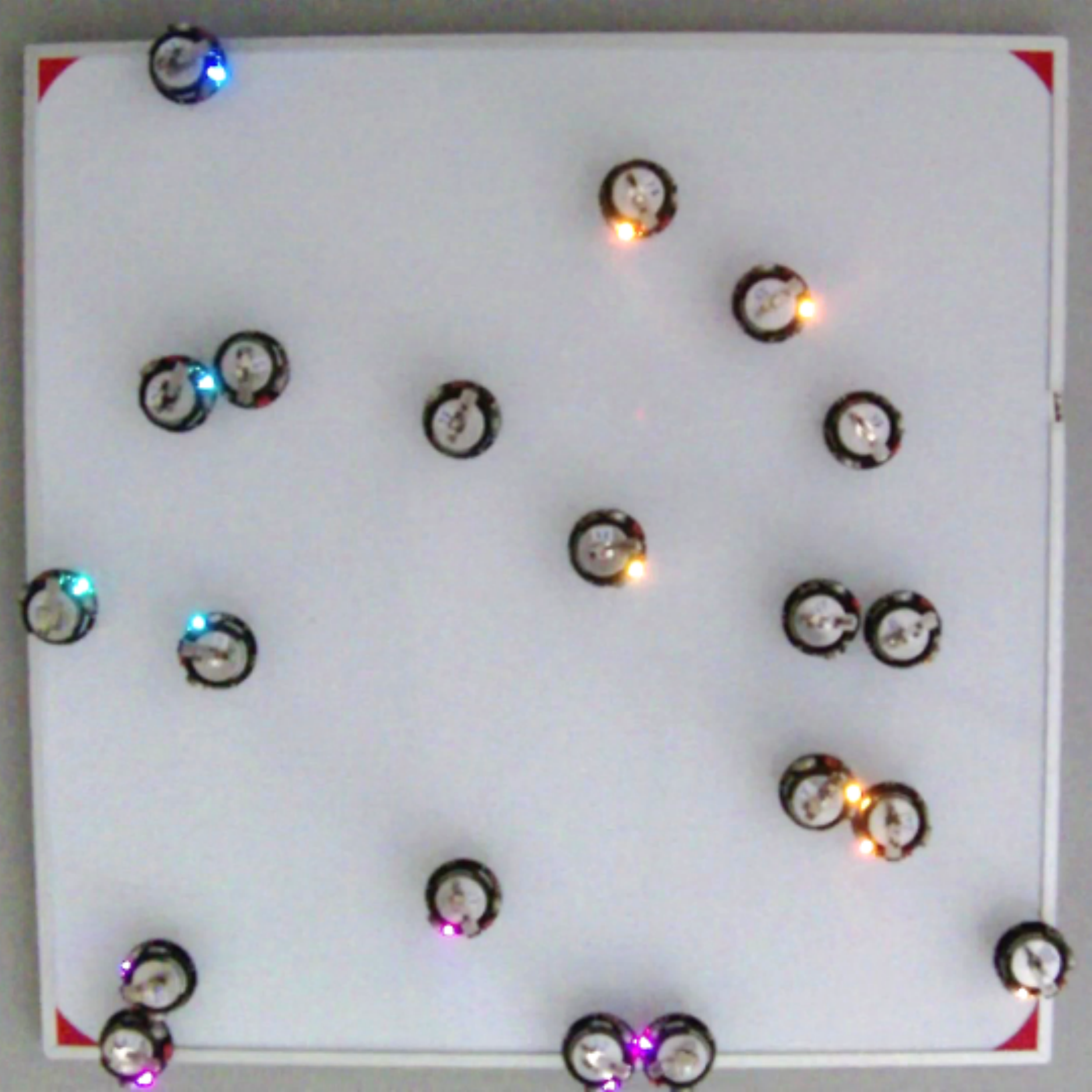}&
    \includegraphics[width=0.18\textwidth]{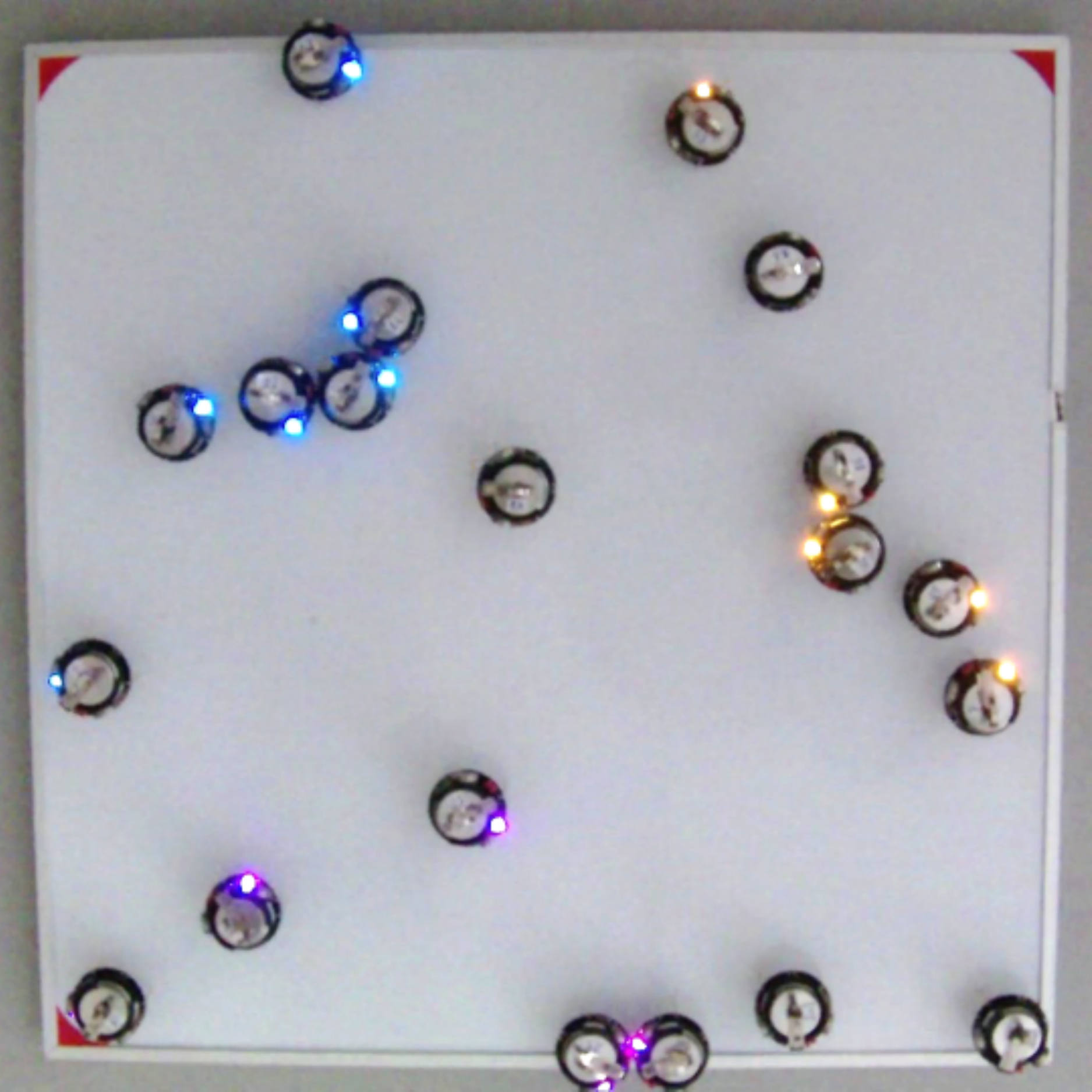}&
    \includegraphics[width=0.18\textwidth]{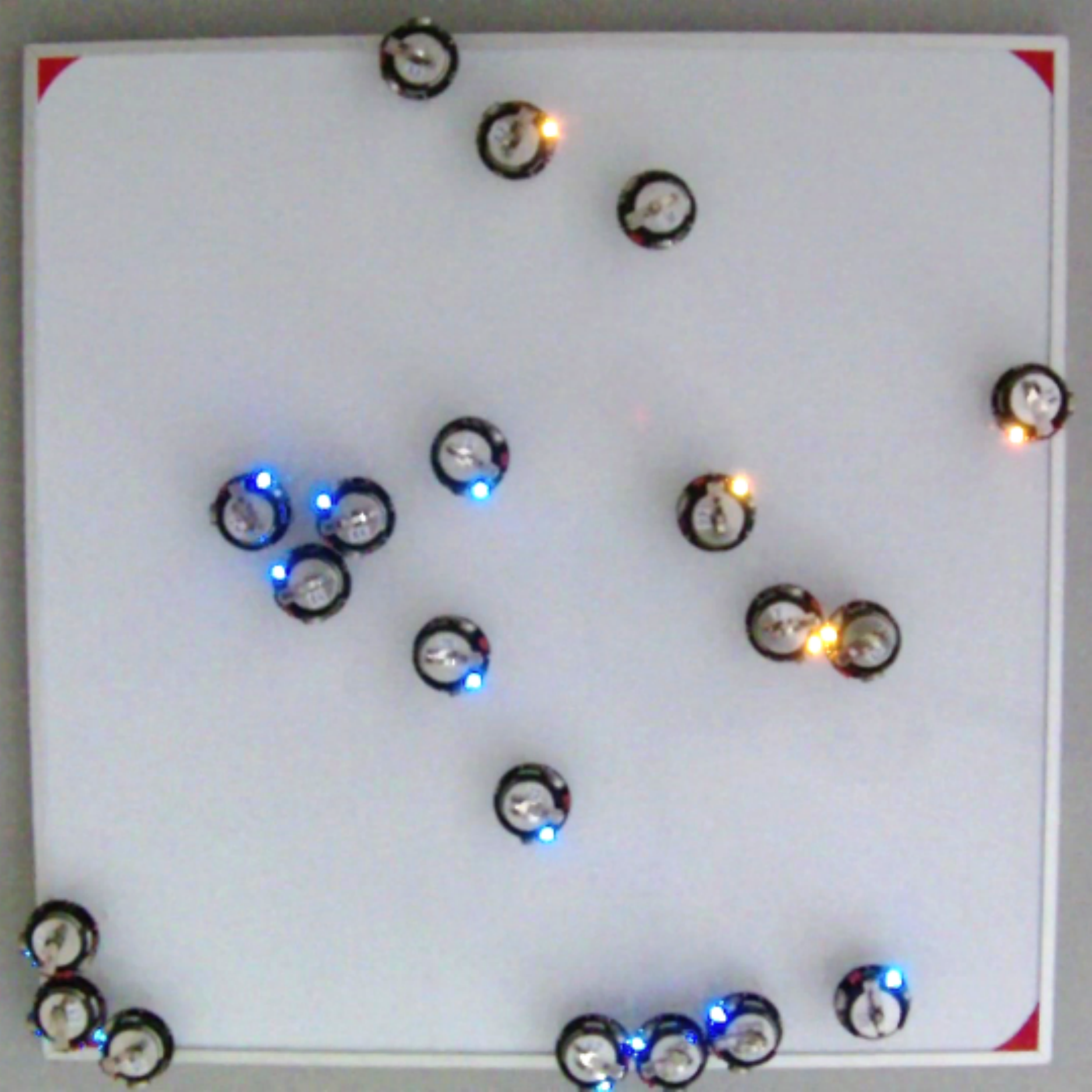}&
    \includegraphics[width=0.18\textwidth]{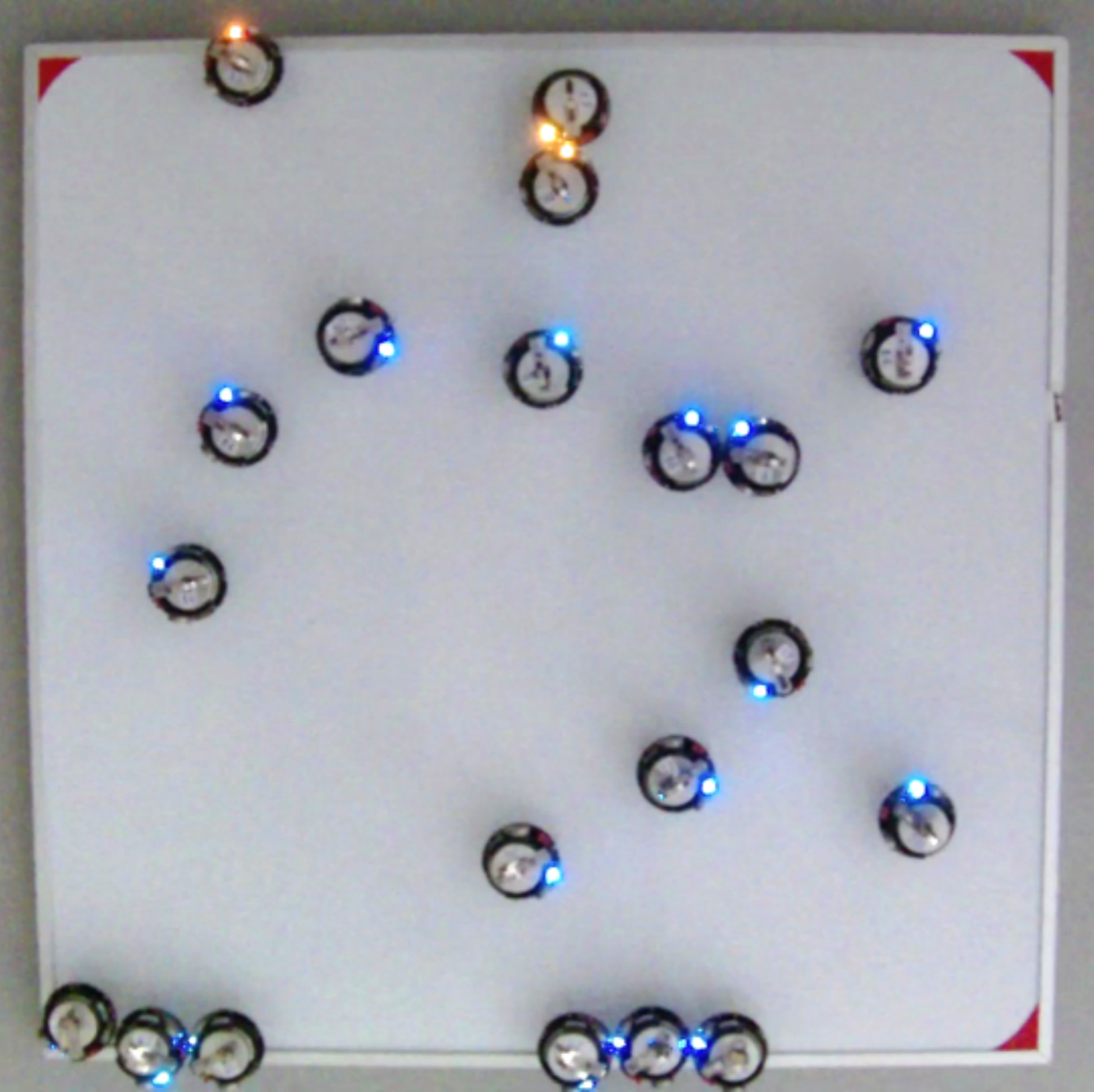}&
    \includegraphics[width=0.18\textwidth]{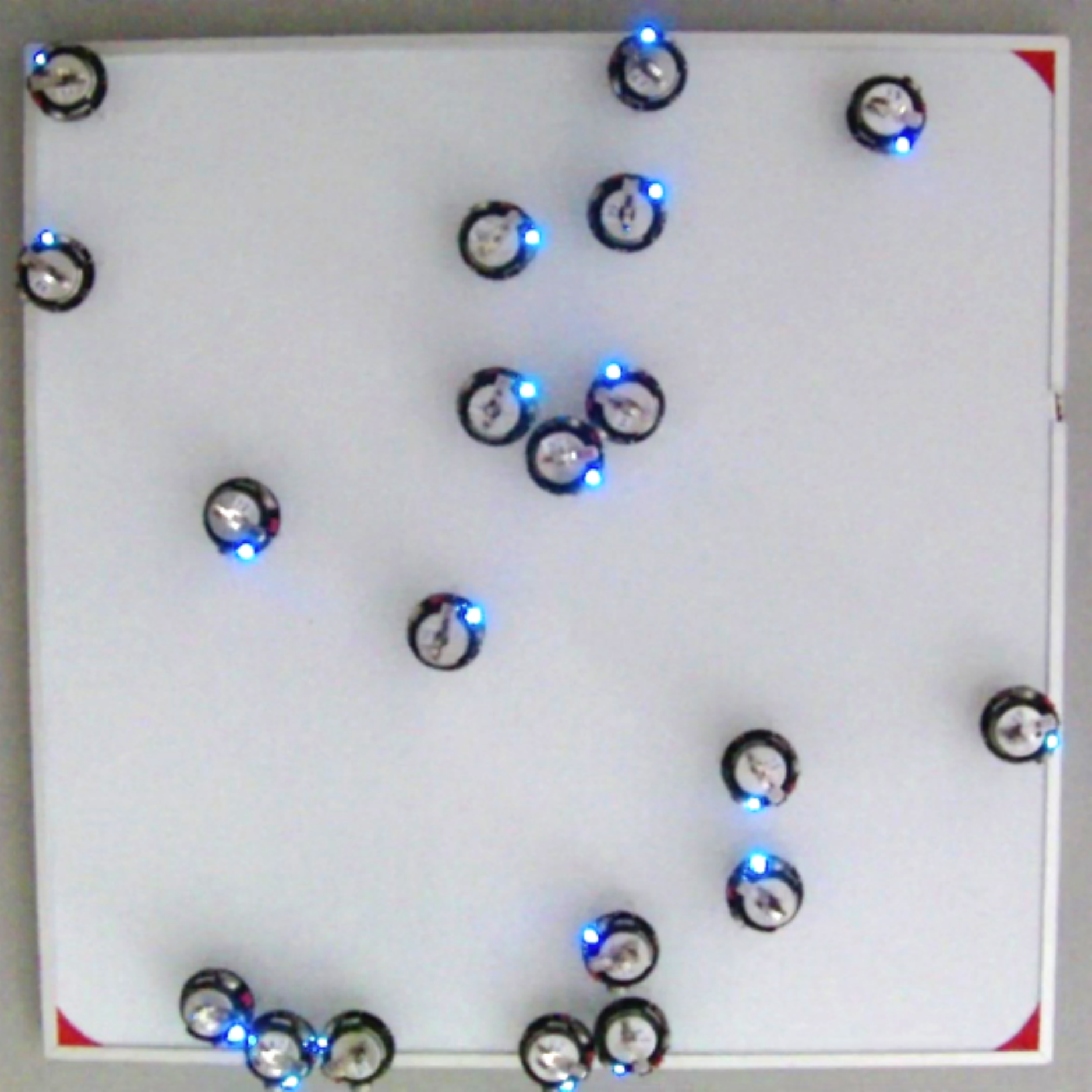}\\
    $t=\unit[10]{s}$&
    $t=\unit[23]{s}$&
    $t=\unit[45]{s}$&
    $t=\unit[60]{s}$&
    $t=\unit[80]{s}$\\
  \end{tabular}
  \caption{Different shots of an experiment with 20 kilobots, with
    $\tau_s = \tau_m = \unit[2.5]{s}$. Robots lighting their LED have
    only one word in their inventory, while no color signal indicates
    more than one word or an empty inventory. Different words
    correspond to different colours.}
  \label{fig:shots}
\end{figure}

\subsection{Experiments with real robots}
\label{sec:validation-with-real}

To validate our results with respect to the real robotic platform, we
performed comparative experiments in a smaller arena
($L=\unit[45]{cm}$). In this condition, we have $N_1\simeq 6$ and
$N_c\simeq 29$, which led us to use smaller groups of robots
($N\in\{5,20,35\}$) to explore the system behaviour as the
characteristics of the static interaction network vary.  Given the
smaller dimensions, we also explored smaller values for the latencies
$\tau_s$ and $\tau_m$, and we decided to set both to the same value
$\tau_a\in\unit[\{2.5,5,7.5\}]{s}$. We have performed 20 runs with
real robots for each of the 9 experimental conditions (3 group sizes
$\times$ 3 latencies), for a total of 180 runs. Figure~\ref{fig:shots}
shows a sequence of frames from one run performed with 20 kilobots. It
is possible to note that initially multiple small clusters are present
in which robots have the same word (here represented by the color of
the onboard LED). As time goes by, clusters disappear and eventually
one single word is chosen. 

\begin{figure}[!t]
  \centering
  \includegraphics[width=0.6\textwidth]{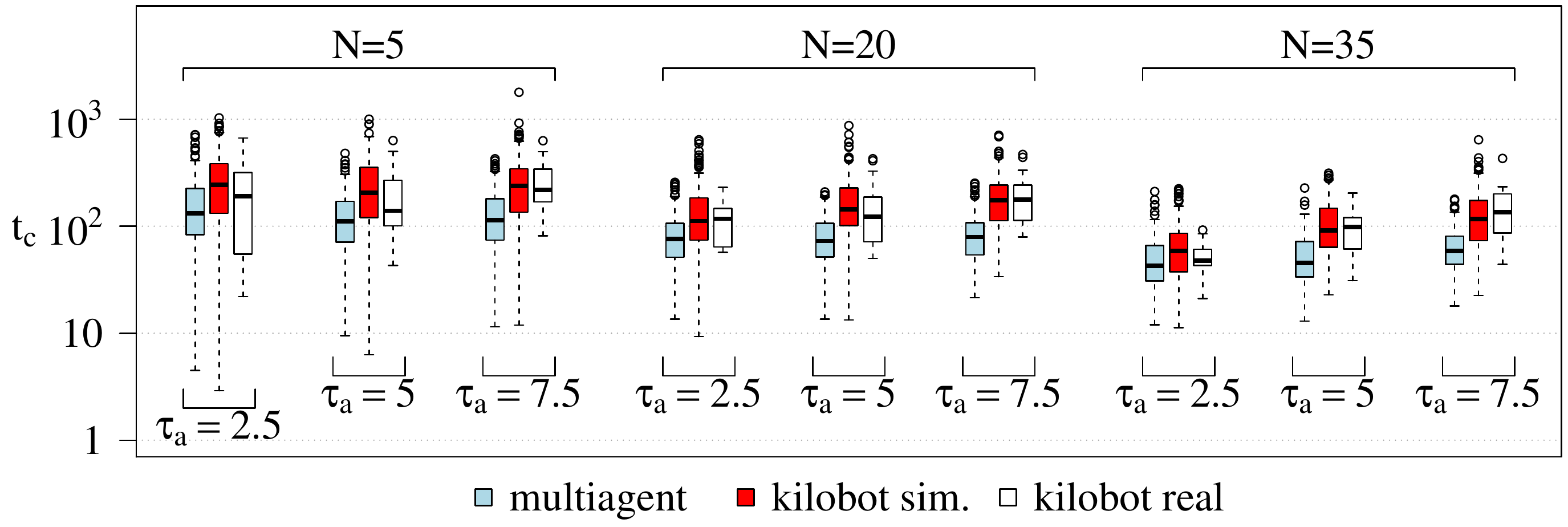}  
  \caption{Comparison between multiagent simulations, kilobot
    simulations and real kilobots. For each condition, 200 runs were
    performed in simulation, while 20 runs were performed with
    physical robots. Boxes represent the interquartile range,
    horizontal lines mark the median, whiskers extend to 1.5 times the
    first quartiles, and dots represent outliers.}
  \label{fig:comparison-real}
\end{figure}

For each experimental run, we have recorded the convergence time
$t_c$, and the obtained results have been contrasted with simulations
in comparable conditions---same arena dimensions $L$ and same
latencies as determined by $\tau_a$---with multiagent and robotic
simulations. Figure~\ref{fig:comparison-real} shows that the
statistics are aligned between kilobot simulations and real kilobots,
and both present a slightly larger convergence time with respect to
multiagent simulations. We also note that in general, the convergence
time for $\tau_a=\unit[2.5]{s}$ is lower in case of real robots than
in simulation, while this is not always true for larger
latencies. This is an effect of interferences in communication due to
simultaneous broadcasts, which leads to the loss of some communication
messages. When messages get lost, the convergence dynamics are
actually faster because the exchanging of different words by robots
broadcasting at the same time gets reduced. This is in line with the
observations made in Section~\ref{sec:consensus-dynamics} about the
influence of the broadcasting period, indicating that convergence is
faster when there are less broadcasts.  With kilobots, a reduction of
the number of broadcasts due to interference results from small
$\tau_s$ and large $N$ (see Fig.~\ref{fig:comparison-real}).

\section{Conclusions}
\label{sec:conclusions}

This study finds itself at the interface between theoretical
investigations and robotics implementation. The results observed in
the multiagent simulations can be of interest for complex systems
studies as they highlight the effects of concurrency and of different
latencies in the motion and interaction patterns, as determined by
implementation constraints. Concurrency is customary in multi-robot
systems and artificial decentralised systems in general. Hence,
accounting for it into abstract models is important to provide usable
predictions. We have found here that concurrent executions of games
are particularly important for aspects like the maximum memory $M$,
and future studies should better characterise such effects in terms of
the probability of observing concurrent executions at any time.

Robotics simulations and experiments with kilobots showed how
embodiment influences the consensus dynamics by limiting the diffusion
of information into the system: on the one hand, collisions lead to
the formation of clusters that dissolve slower for larger $\tau_m$,
leading to slower convergence times. On the other hand, the memory
requirements of robots is reduced as only few robots at the interface
between clusters experience more than two words at the same time.
Future studies should attempt a more precise description of the
diffusive motion of agents under physical constraints, in order to
obtain better predictions in terms of the expected interaction
network. Additionally, the communication protocol employed by kilobots
and the observed interferences need to be better
characterised. Simulations should account for uncertain reception of
messages, as well as for the exponential back-off used during
transmission when the channel is busy. By including such features, we
expect to deliver precise estimations of the system behaviour even for
very large group sizes and short broadcasting periods.

%
\bibliographystyle{plain}
\bibliography{trianni-etal} 

\end{document}